\documentclass[aps,pra,twocolumn,superscriptaddress,showpacs,showkeys,amsmath,amssymb]{revtex4-1}

\usepackage{amsmath}
\usepackage{accents}
\usepackage{bm}
\usepackage[english]{babel}
\usepackage{graphicx}
\usepackage[usenames]{color}
\usepackage{colortbl}
\usepackage[font=footnotesize,figurewithin=none]{caption}
\usepackage{subcaption}
\usepackage{floatrow}
\usepackage[toc,page,title]{appendix}
\usepackage{CJK}

\begin{document}

\begin{CJK*}{GB}{}

\title{Detecting entanglement by the mean value of spin on a quantum computer}

\author{A.R.~Kuzmak}
\email{andrijkuzmak@gmail.com}
\author{V.M.~Tkachuk}
\email{voltkachuk@gmail.com}
\affiliation{Department for Theoretical Physics, Ivan Franko National University of Lviv, 12 Drahomanov St., Lviv, UA-79005, Ukraine}

\date{\today}
\begin{abstract}
We implement a protocol to determine the degree of entanglement between a qubit and the rest of the system on a quantum computer. The protocol is based on results obtained in paper [A. M. Frydryszak, M. I. Samar, V. M. Tkachuk, Eur. Phys. J. D {\bf 71} 233 (2017)].
This protocol is tested on a 5-qubit superconducting quantum processor called ibmq-ourense provided by the IBM company.
We determine the values of entanglement of the Schr\"odinger cat and the Werner states prepared on this device and compare them with
the theoretical ones. In addition, a protocol for determining the entanglement of rank-2 mixed states is proposed. We apply this protocol
to the mixed state which consists of two Bell states prepared on the ibmq-ourense quantum device.
\end{abstract}

\maketitle

\end{CJK*}

\section{Introduction \label{sec1}}

Entanglement is an inherent property of a quantum system \cite{EPRP,horodecki2009}. It plays a crucial role in processes related to quantum information
and quantum computation \cite{feynman1982,nielsen2000}. The presence of entanglement in a system allows one to implement various quantum information schemes and devices
that cannot be realized by classical systems. The application of this phenomenon to the implementation of various quantum algorithms
began after Aspect et al. \cite{ASPECT} tested Bell's inequality \cite{BELL} and experimentally solved the EPR paradox \cite{EPRP}.
The preparation of an entangled state is an indispensable step in the realization of such tasks as quantum cryptography \cite{Ekert1991}, super-dense coding
\cite{Bennett1992}, teleportation \cite{TELEPORT}, etc.

Quantum entanglement is a key resource that is crucial in the efficient and fast modeling of many-body quantum systems on
quantum computers \cite{feynman1982,lloyd1996,buluta2009,preskill2012}. Due to this feature, quantum computers are much more efficient than classical ones
for studying different problems relating to the behavior of quantum systems in various fields, including condensed-matter physics, high-energy physics, atomic physics, and quantum chemistry.

In recent years, physical implementation of quantum computers on superconducting circuits have achieved significant progress.
The IBM company has developed a cloud service called IBM Q Experience \cite{IBMQExp}. It has access to different quantum devices
based on processors containing from 1 up to 20 superconducting qubits. Recently, it was shown that the 16-qubit \cite{wang2018s}
and 20-qubit \cite{mooney2019} IBM Q quantum processors can be fully entangled. Authors of the papers made the quantum tomography on each pair of connected qubits
prepared in the graph states, then calculated the negativity as a measure of entanglement between them. They obtain that the state
is inseparable with respect to any fixed pair. The entangled state were also prepared and measured on other systems with full qubit control,
namely, 20-qubit ion trap system \cite{monz2011,friis2018}, system of photons \cite{wang2016,wang2018,zhong2018} and superconducting
system \cite{song2017,gong2019}.

In this paper, we implement the protocol for measuring the entanglement of one qubit with the rest of the system on a quantum
computers. This protocol is based on results obtained in paper \cite{frydryszak2017}. For this purpose, in the case of pure a state, we use the definition of the geometric measure of entanglement by the mean value of spin proposed in paper \cite{frydryszak2017}. We test this protocol on the ibmq-ourense quantum device \cite{IBMQExp}.
Therefore, the entanglement of the Schr\"odinger cat and Werner states are defined. In addition, we propose and test a protocol to determine
the entanglement of rank-2 mixed quantum states.

\section{Protocol for measuring the entanglement of a pure state by the mean value of spin \label{sec2}}

\begin{figure}[!!h]
\includegraphics[scale=0.4500, angle=0.0, clip]{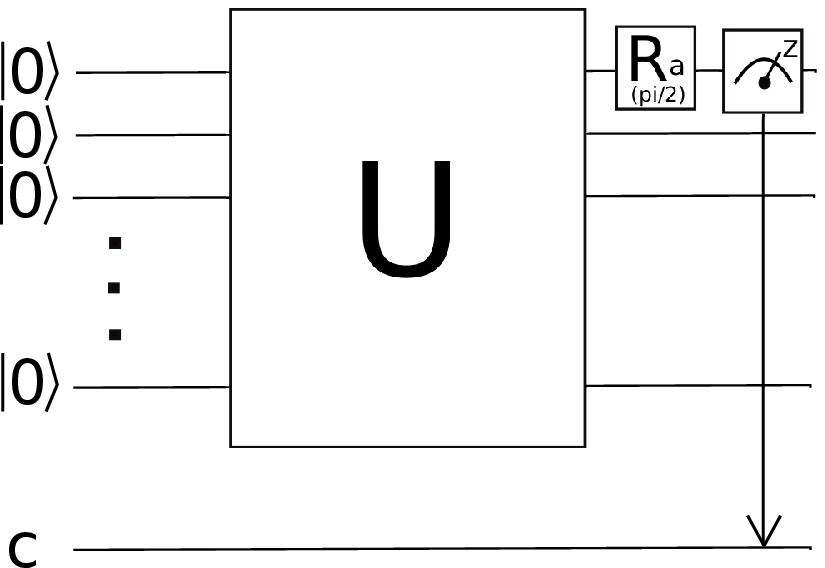}
\caption{Quantum circuit for measuring the mean value of Pauli $a$-operator of the first spin in the state generated
by the unitary operator $U$. The unitary operator $U$ transforms the system from the initial state to state (\ref{generalpureqs}).
The $R_a$ gate provides the rotation of the qubit state around the $a$-axis by angle $\pi/2$.}
\label{mvpo}
\end{figure}

We propose a method which allows one to measure the entanglement of the state prepared on quantum computer. This method bases on the definition of the geometric measure of entanglement by the mean value of spin proposed in paper \cite{frydryszak2017}. If we have a spin that can be entangled with the rest of the system in a pure state
\begin{eqnarray}
\vert\psi\rangle=a\vert 0\rangle\vert\phi_1\rangle+b\vert 1\rangle\vert\phi_2\rangle
\label{generalpureqs}
\end{eqnarray}
then the degree of entanglement between these subsystems can be defined by the mean value of the spin as follows
\begin{eqnarray}
E\left(\vert\psi\rangle\right)=\frac{1}{2}\left(1-\vert\langle\psi\vert{\bm \sigma}\vert\psi\rangle\vert\right),
\label{geommeasure}
\end{eqnarray}
where $a$ and $b$ are some complex constants which satisfy the normalization condition $\vert a\vert^2+\vert b\vert^2=1$;
$\vert\phi_1\rangle$ and $\vert\phi_2\rangle$ are the state vectors which define the quantum system entangled with the spin, and
which satisfy the normalization conditions $\langle\phi_1\vert\phi_1\rangle=1$, $\langle\phi_2\vert\phi_2\rangle=1$;
modulus of the mean value of the spin is determined by the expression $\vert\langle\psi\vert{\bm \sigma}\vert\psi\rangle\vert=\sqrt{\langle\psi\vert{\bm \sigma}\vert\psi\rangle^2}$, and
operator ${\bm \sigma}$ is defined by the Pauli matrices as follows ${\bm \sigma}={\bf i}\sigma^x+{\bf j}\sigma^y+{\bf k}\sigma^z$.
Note that, in general case, functions $\vert\phi_1\rangle$ and $\vert\phi_2\rangle$ are not orthogonal, i.e. $\langle\phi_1\vert\phi_2\rangle\neq 0$.
Since any two-level quantum system is described by the Pauli matrices, expression (\ref{geommeasure}) can be applied to the determination of the value of entanglement of any quantum system that consists of a set of two-level subsystems.
Thus, to determine the entanglement of one qubit with the rest of the system the mean value of the Pauli operator corresponding to this qubit
in state (\ref{generalpureqs}) should be measured. It is convenient to use basis $\vert 0\rangle$, $\vert 1\rangle$
for measuring certain value of Pauli operator. For instance, the tomography process of the IBM quantum computers is provided
by the measurements of a state on this basis. For this purpose, we represent the mean values of the spin
by the values which can be measured using basis $\vert 0\rangle$, $\vert 1\rangle$. The fact that the states $\vert 0\rangle$, $\vert 1\rangle$
are the eigenstates of the $\sigma^z$ operator with $\pm 1$ eigenvalues, respectively, allows us to decompose it as follows
$\sigma^z=\vert 0\rangle\langle 0\vert-\vert 1\rangle\langle 1\vert$.
In turn, this allows us to express the mean value of this operator in state (\ref{generalpureqs}) by the probabilities which define the result of measure of qubit on basis $\vert 0\rangle$, $\vert 1\rangle$ in the form
\begin{eqnarray}
\langle\psi\vert\sigma^z\vert\psi\rangle=\vert\langle\psi\vert 0\rangle\vert^2-\vert\langle\psi\vert 1\rangle\vert^2 .
\label{mvzpauli}
\end{eqnarray}
To measure the remaining mean values of the spin in basis $\vert 0\rangle$, $\vert 1\rangle$, we represent them as follows
$\sigma^x=e^{-i\frac{\pi}{4}\sigma^y}\sigma^ze^{i\frac{\pi}{4}\sigma^y}$, $\sigma^y=e^{i\frac{\pi}{4}\sigma^x}\sigma^ze^{-i\frac{\pi}{4}\sigma^x}$.
Then the mean value of the $x$-component of the spin takes the form
\begin{eqnarray}
&&\langle\psi\vert\sigma^x\vert\psi\rangle=\langle\psi\vert e^{-i\frac{\pi}{4}\sigma^y}\sigma^ze^{i\frac{\pi}{4}\sigma^y}\vert\psi\rangle=\langle\tilde{\psi}^y\vert\sigma^z\vert\tilde{\psi}^y\rangle\nonumber\\
&&=\langle\tilde{\psi}^y\vert 0\rangle\langle 0\vert\tilde{\psi}^y\rangle-\langle\tilde{\psi}^y\vert 1\rangle\langle 1\vert\tilde{\psi}^y\rangle\nonumber\\
&&=\vert\langle\tilde{\psi}^y\vert 0\rangle\vert^2-\vert\langle\tilde{\psi}^y\vert 1\rangle\vert^2,
\label{mvpaulix}
\end{eqnarray}
where $\vert\tilde{\psi}^y\rangle=e^{i\frac{\pi}{4}\sigma^y}\vert\psi\rangle$. In a similar way we represent the expression for the
mean value of the $y$-component of the spin
\begin{eqnarray}
\langle\psi\vert\sigma^y\vert\psi\rangle=\vert\langle\tilde{\psi}^x\vert 0\rangle\vert^2-\vert\langle\tilde{\psi}^x\vert 1\rangle\vert^2,
\label{mvpauliy}
\end{eqnarray}
where $\vert\tilde{\psi}^x\rangle=e^{-i\frac{\pi}{4}\sigma^x}\vert\psi\rangle$.
As we can see, before measuring the mean value of the $x$- and $y$-component of the spin, it should be rotated around the $y$- and $x$-axis, respectively,
by angle $\pi/2$. Due to the modules in expressions (\ref{mvpaulix}) and (\ref{mvpauliy}), the directions of rotation are not important.
The protocol to determine the mean values of the Pauli operators of the first spin is presented in the Fig.~\ref{mvpo}. Note that
this protocol is valid for determining the mean values of Pauli operators of any spins. Let us apply this protocol for measuring the entanglement
of quantum states prepared on the ibmq-ourense quantum computer.

\section{Measuring the entanglement of pure states prepared on the ibmq-ourense quantum computer \label{sec3}}

\begin{figure}[!!h]
\includegraphics[scale=0.400, angle=0.0, clip]{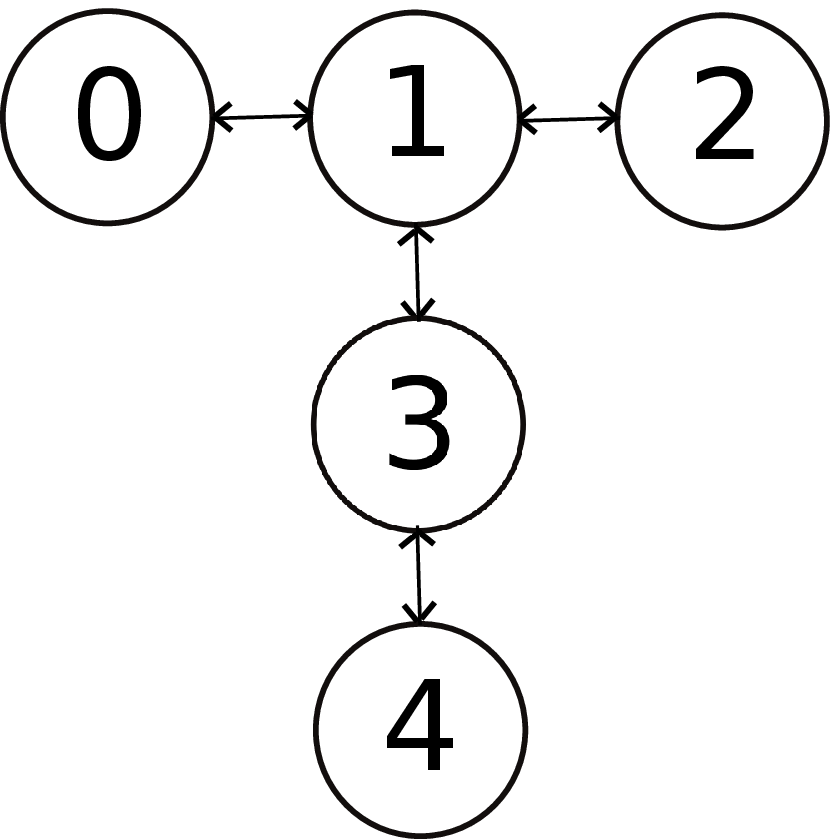}
\caption{Structure of the ibmq-ourense quantum device. This device consists of five superconducting qubits which interact between themselves as are
shown in the figure. The arrows connect the qubits between which the controlled-NOT gate can be directly applied. Bidirectionality of arrows means that each of the qubit can be both a control and a target.}
\label{ibmq-ourense}
\end{figure}

In this section we use the protocol proposed above to measure the entanglement of states prepared on the ibmq-ourence quantum computer. This is one of the quantum
devices to which the IBM provides free access via its cloud service. This device consists of five superconducting qubits in the way shown in Fig.~\ref{ibmq-ourense}.
We prepare pure states, namely, the Schr\"odinger cat states and the Werner-like states, and apply to them the protocol considered in the previous
section.

\subsection{Schr\"odinger cat states \label{subsec3_1}}

\begin{figure}[!!h]
\includegraphics[scale=0.500, angle=0.0, clip]{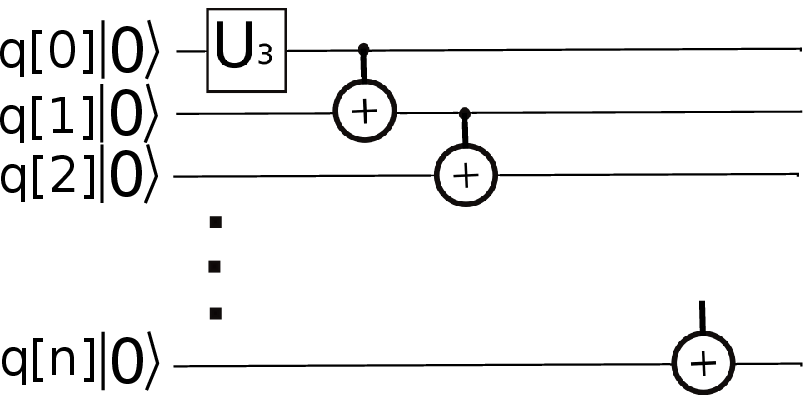}
\caption{The circuits for the preparation of a general Schr\"odinger cat state on a system of $n$ qubits.}
\label{HCatstate}
\end{figure}

In general, the Schr\"odinger cat state for $n$ qubits reads
\begin{eqnarray}
\vert\psi_{cat}\rangle =\cos\frac{\theta}{2}\vert 00\ldots 0\rangle + \sin\frac{\theta}{2}e^{i\phi}\vert 11\ldots 1\rangle,
\label{Heisenbergstate}
\end{eqnarray}
where $\theta\in[0,\pi]$ and $\phi\in[0,2\pi]$ are some real parameters. This state can be prepared by applying the single-qubit gate $U_3(\theta,\phi,\lambda)$ to the initial state
$\vert 00\ldots 0\rangle$ and a sequence of controlled-NOT operators as shown in Fig.~\ref{HCatstate}.
In the basis $\vert 0\rangle$, $\vert 1\rangle$ the matrix representation of $U_3(\theta,\phi,\lambda)$ gate has the following form
\begin{eqnarray}
U_3(\theta,\phi,\lambda)=\left( \begin{array}{ccccc}
\cos\frac{\theta}{2} & -e^{i\lambda}\sin{\frac{\theta} {2}} \\
e^{i\phi}\sin{\frac{\theta} {2}} & e^{i\left(\lambda+\phi\right)}\cos\frac{\theta}{2}
\end{array}\right),
\label{U3gate}
\end{eqnarray}
where $\lambda$ is a real parameter which can take the values from the range $[0,2\pi]$. This gate has the effect of rotating a qubit
in the initial state $\vert 0\rangle$ to an arbitrary one-qubit state
\begin{eqnarray}
U_3(\theta,\phi,\lambda)\vert 0\rangle =\cos\frac{\theta}{2}\vert 0\rangle+\sin\frac{\theta}{2}e^{i\phi}\vert 1\rangle.
\label{arboneqstate}
\end{eqnarray}

\begin{figure}[!!h]
\includegraphics[scale=0.57, angle=0.0, clip]{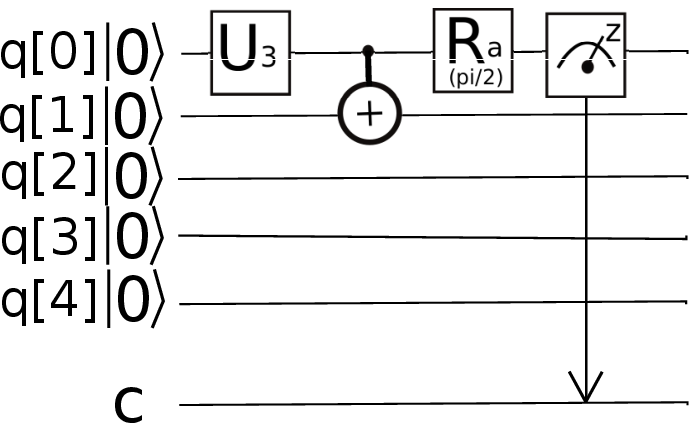}
\includegraphics[scale=0.57, angle=0.0, clip]{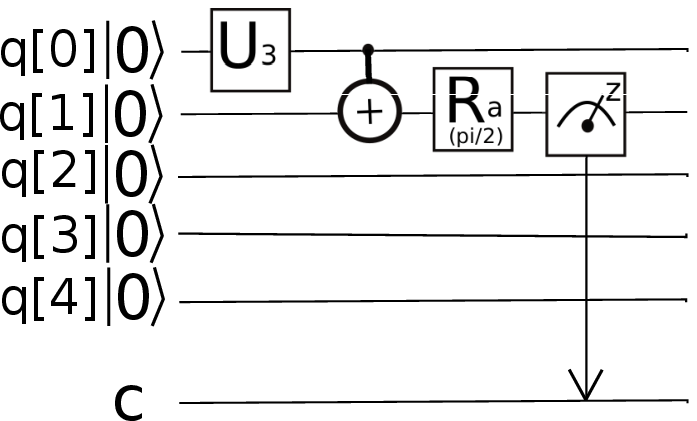}\\
\subcaptionbox{\label{Hcatstate_2qubit_m0q}}{\includegraphics[scale=0.30, angle=0.0, clip]{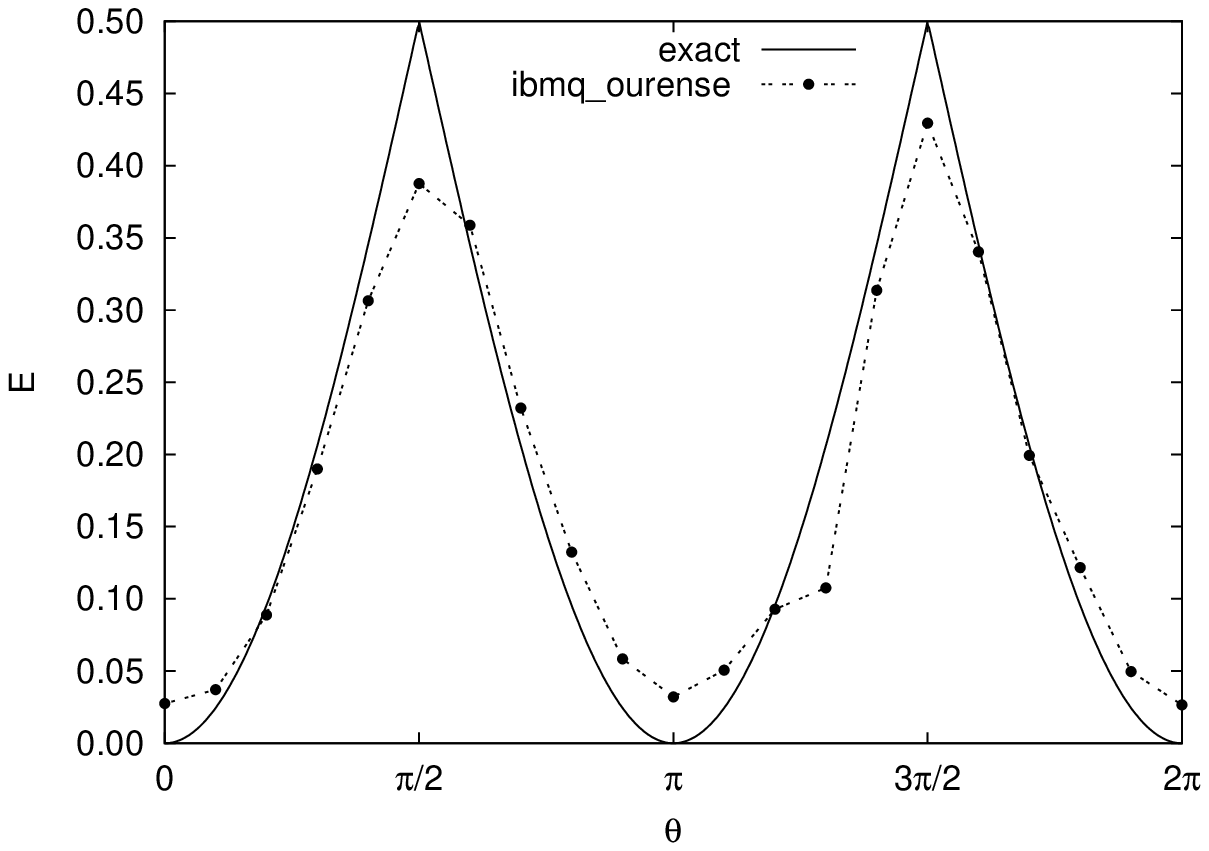}}
\subcaptionbox{\label{Hcatstate_2qubit_m1q}}{\includegraphics[scale=0.30, angle=0.0, clip]{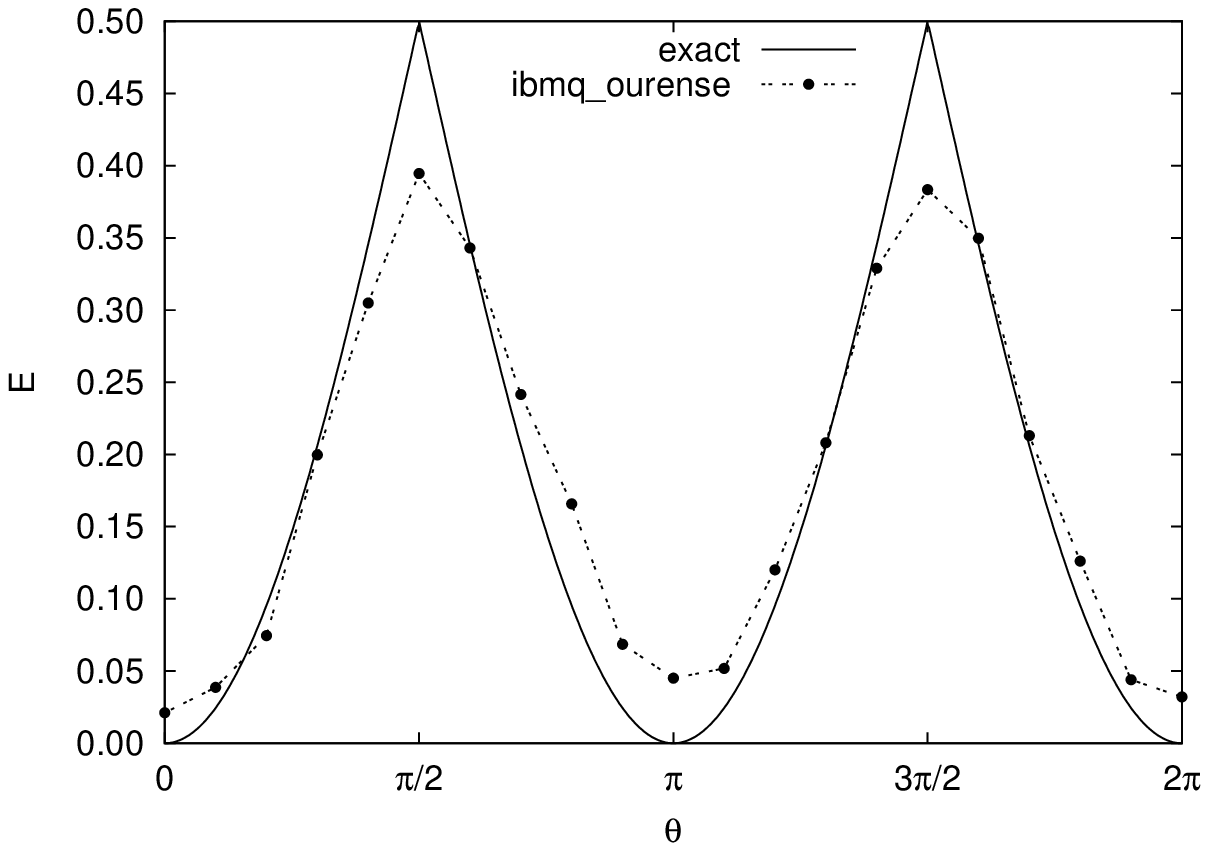}}
\caption{Circuits for preparation and measurement of the two-qubit Schr\"odinger cat state and corresponding dependences of entanglement
in the case of measuring of first q[0] (a) and second q[1] (b) qubits, respectively. Here operators $R_a$ provide the rotations of the
qubit state around the $a=x,y,z$ axises by the angle $\pi/2$.}
\label{Hcatstate_2qubit}
\end{figure}

The entanglement of any qubit with the remaining system in state (\ref{Heisenbergstate}) according to (\ref{geommeasure}) is
\begin{eqnarray}
E\left(\vert\psi_{cat}\rangle\right)=\frac{1}{2}\left(1-\vert\cos\theta\vert\right).
\label{geommeasurecat}
\end{eqnarray}
On the ibmq-ourense quantum device we measure the dependence of the geometry measure of entanglement on the parameter $\theta$
for two-qubit Schr\"odinger cat state with $\phi=0$. The results are obtained for two qubits (Fig.~\ref{Hcatstate_2qubit}). Figs.~\ref{Hcatstate_2qubit_m0q}
and \ref{Hcatstate_2qubit_m1q} show the circuits and the dependence of entanglement in the case of measuring first q[0] and second q[1] qubits, respectively. These results are almost identical. This is caused by the fact that the difference between single-qubit errors is compensated by the difference between the readout errors of each qubit. Indeed, the error of the q[0] qubit, $4.18\times 10^{-4}$, is greater than the q[1], $3.88\times 10^{-4}$, and vice versa the readout error of the q[0] qubitm, $1.90\times 10^{-2}$, is less than the readout error of the q[1] qubit, $3.70\times 10^{-2}$.
In addition, the coherence time of the q[0] qubit, $T_2=73.01$ ${\rm \mu}$s, is almost two times longer than the coherence time of the q[1] qubit, $T_2=36.30$ ${\rm \mu}$s. The difference between single-qubit errors is also compensated by the difference between the readout errors. It is important to note that the number of measurements of a pure quantum state, which we considered in this paper, is equal to 1024. Thus, a quantum computer makes 1024 shots to obtain the mean value of the $\alpha$-component of certain spin in a predefined pure quantum state. After calculation on quantum computer we obtain the probabilities to find this spin in states $\vert 0\rangle$ and $\vert 1\rangle$. In order to obtain the mean value of spin components, we substitute these probabilities into equations (\ref{mvzpauli})--(\ref{mvpauliy}). Then, to obtain the value of entanglement these mean values, we substitute into equation (\ref{geommeasure}). Note that the more shots are made by the quantum computer, the smaller is the error due to counting statistics, which is inversely proportional to the square root of the shots (in our case $\sim 1/32$).

\begin{figure}[!!h]
\includegraphics[scale=0.50, angle=0.0, clip]{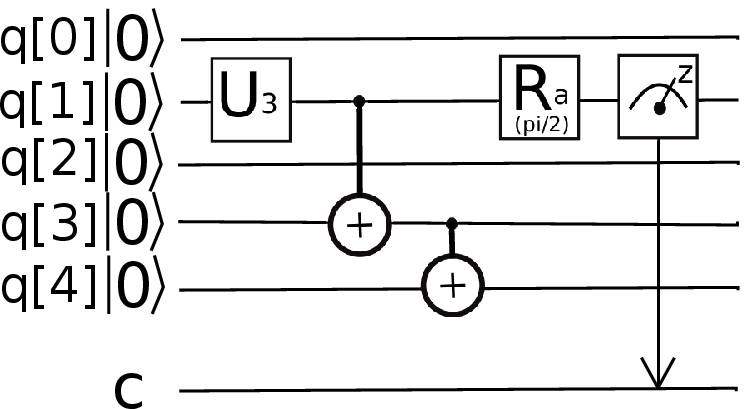}
\includegraphics[scale=0.50, angle=0.0, clip]{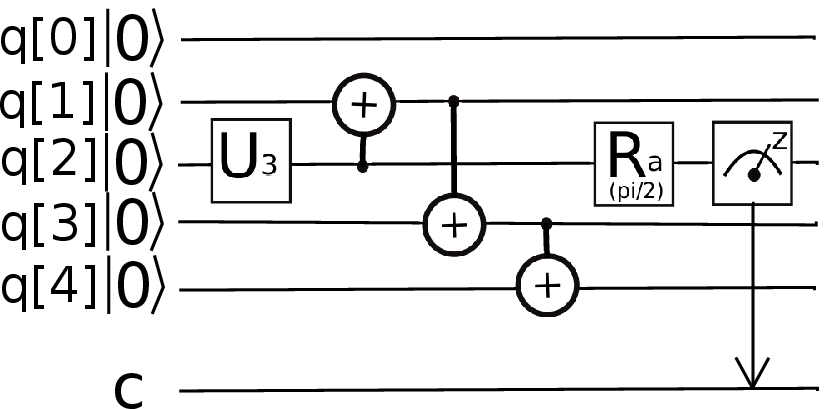}\\
\subcaptionbox{\label{Hcatstate_3qubit}}{\includegraphics[scale=0.33, angle=0.0, clip]{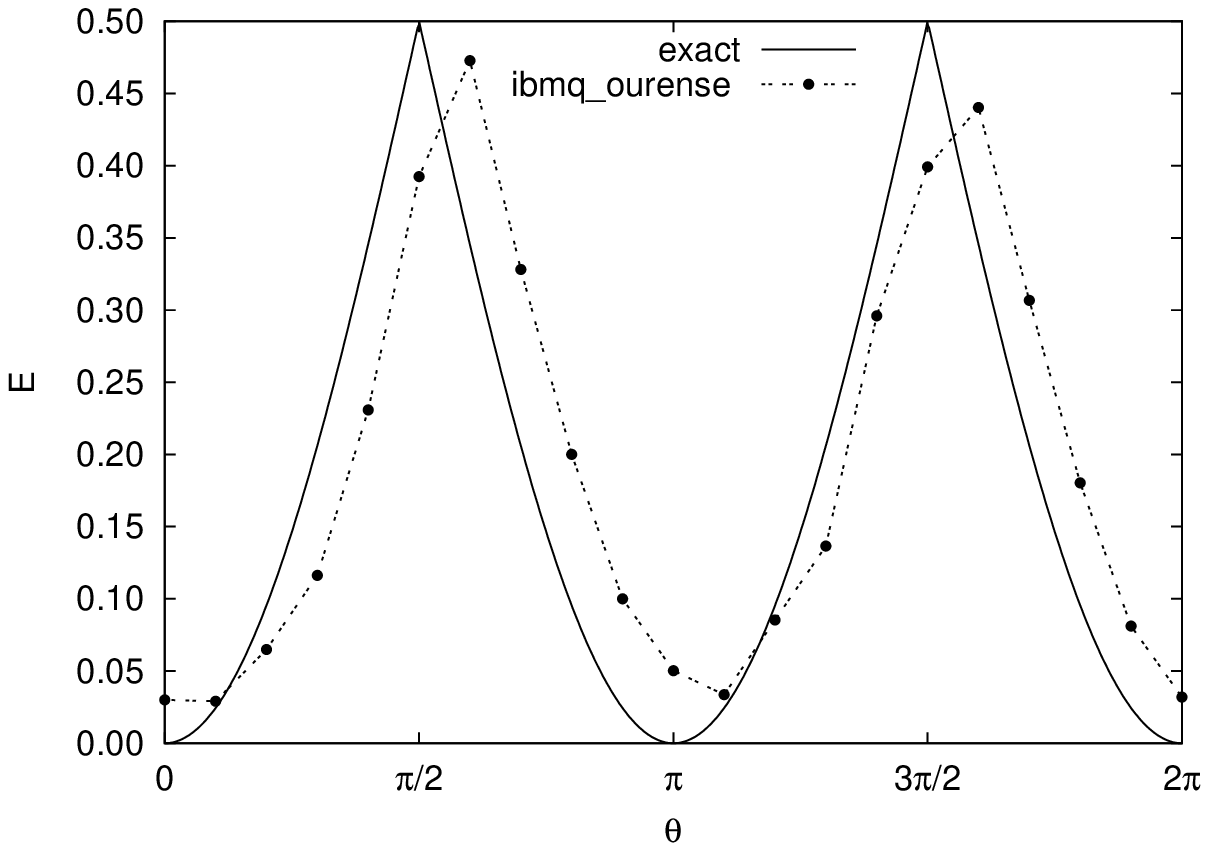}}
\subcaptionbox{\label{Hcatstate_4qubit}}{\includegraphics[scale=0.33, angle=0.0, clip]{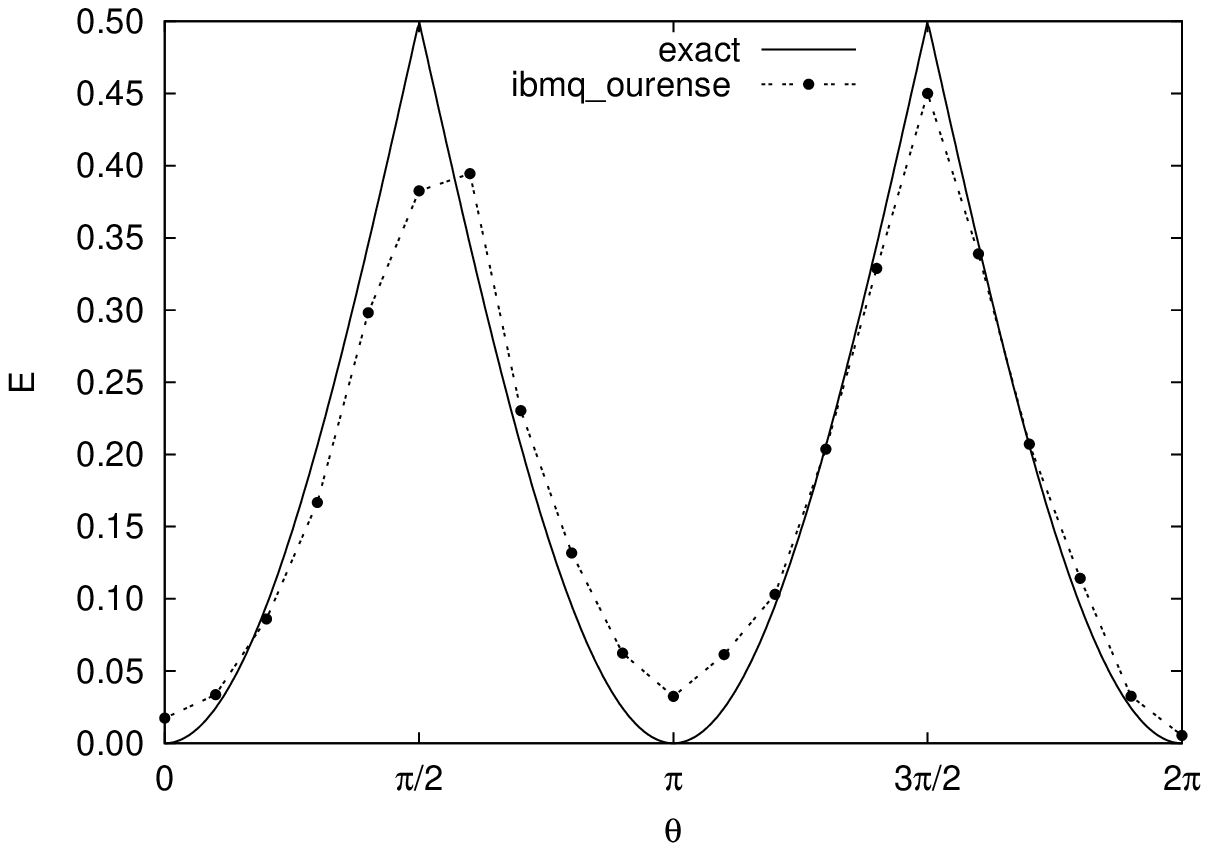}}
\caption{Circuits for preparation and measurement of the Schr\"odinger cat state on the 3 (a) and 4 (b) qubits and corresponding dependences of entanglement. Here operators $R_a$ provide the rotations of the qubit state by the angle $\pi/2$ around the $a=x,y,z$ axises.}
\label{Hcatstate_34qubit}
\end{figure}

Also we prepare and measure the entanglement of the 3- and 4-qubit Schr\"odinger cat states (Fig.~\ref{Hcatstate_34qubit}). Despite the fact that
in these cases we have more qubits in the system, the results obtained by the quantum computer are in good agreement with the theoretical predictions.
This is because the gates that generate the Schr\"odinger cat state are basis operators of the ibmq-ourense quantum device. This fact allows us to reduce the error of preparation of this state. Also in Fig.~\ref{Hcatstate_3qubit} one can see the deviation of the experimental curve from the theoretical one. The precise reason for these deviation is unclear, but it can be due to some systematic errors of the quantum device.

\subsection{Werner state \label{subsec3_2}}

\begin{figure}[!!h]
\includegraphics[scale=0.500, angle=0.0, clip]{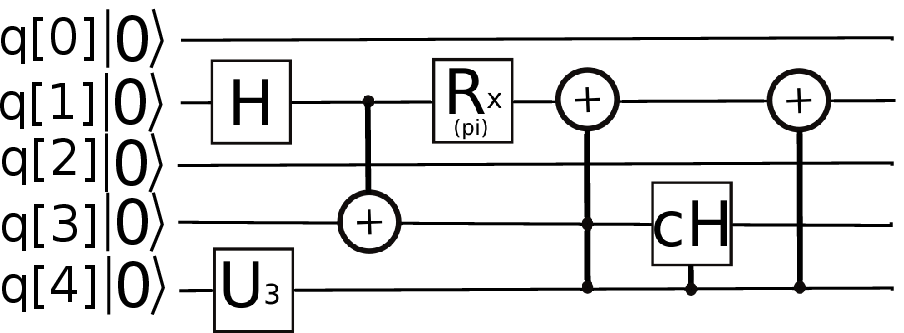}
\caption{Circuit for preparation of the three-qubit Werner state.}
\label{Wstate}
\end{figure}

So far we have studied the entanglement of symmetrical states in the context that the measure of entanglement by the mean value of spin
are the same with respect to any qubit from the system. Now we consider the entanglement of the Werner-like state
\begin{eqnarray}
\vert\psi_{W}\rangle=\sin\frac{\theta}{2}\vert 001\rangle+\frac{1}{\sqrt{2}}\cos\frac{\theta}{2}\left(\vert 010\rangle+\vert 100\rangle\right) ,
\label{Wernerstate}
\end{eqnarray}
which is not symmetric. In order to prepare this state, one should use a set of  additional gates. Thus, for this purpose we use the Hadamard gate ($H$), the $R_x(\pi)$ gate, which provides a single-qubit rotation around the $x$-axis by the angle $\pi$, the controlled-Hadamard gate ($cH$) which
performs an $H$ on the target qubit whenever the control qubit is in state $\vert 1\rangle$, and the Toffoli gate. The circuit of
the preparation of state (\ref{Wernerstate}) on the q[1], q[3] and q[4] qubits is presented in Fig.~\ref{Wstate}.

\begin{figure}[!!h]
\subcaptionbox{\label{EntWq1}}{\includegraphics[scale=0.30, angle=0.0, clip]{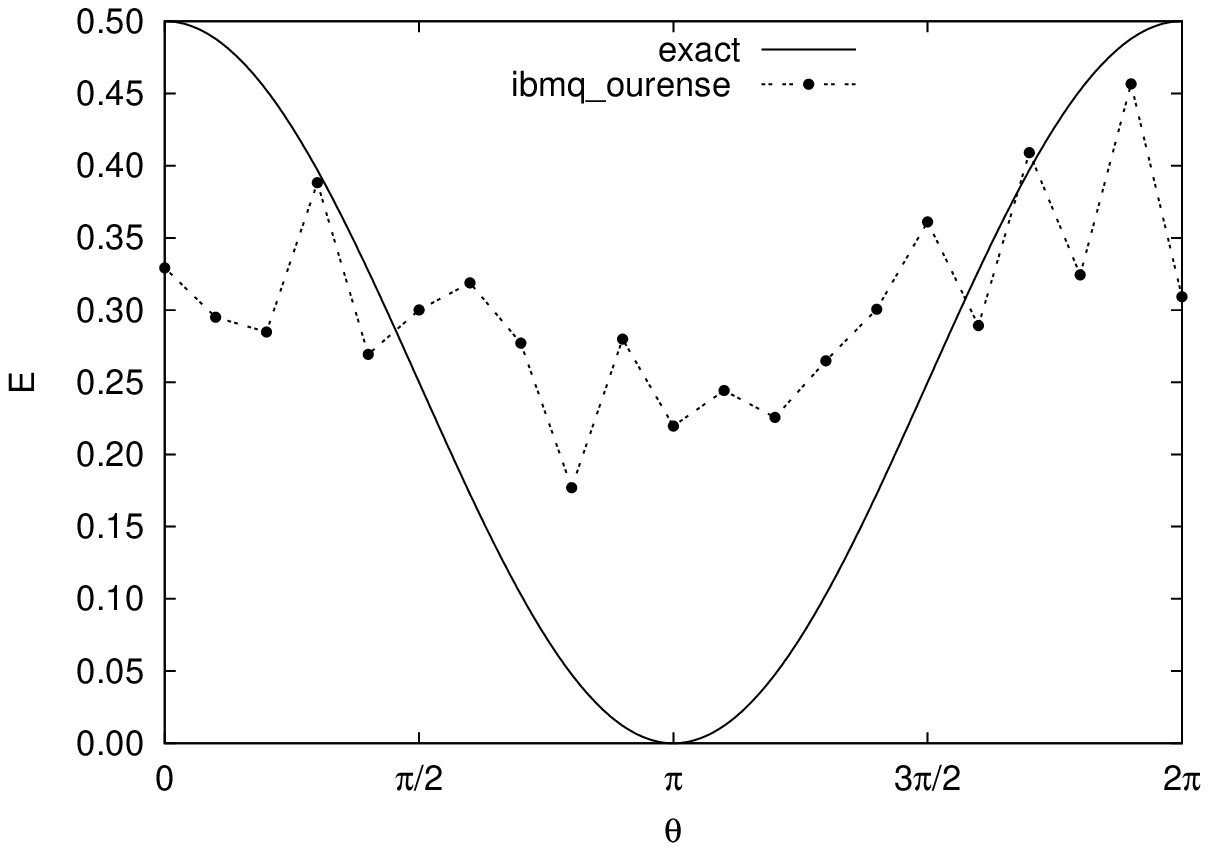}}
\subcaptionbox{\label{EntWq3}}{\includegraphics[scale=0.30, angle=0.0, clip]{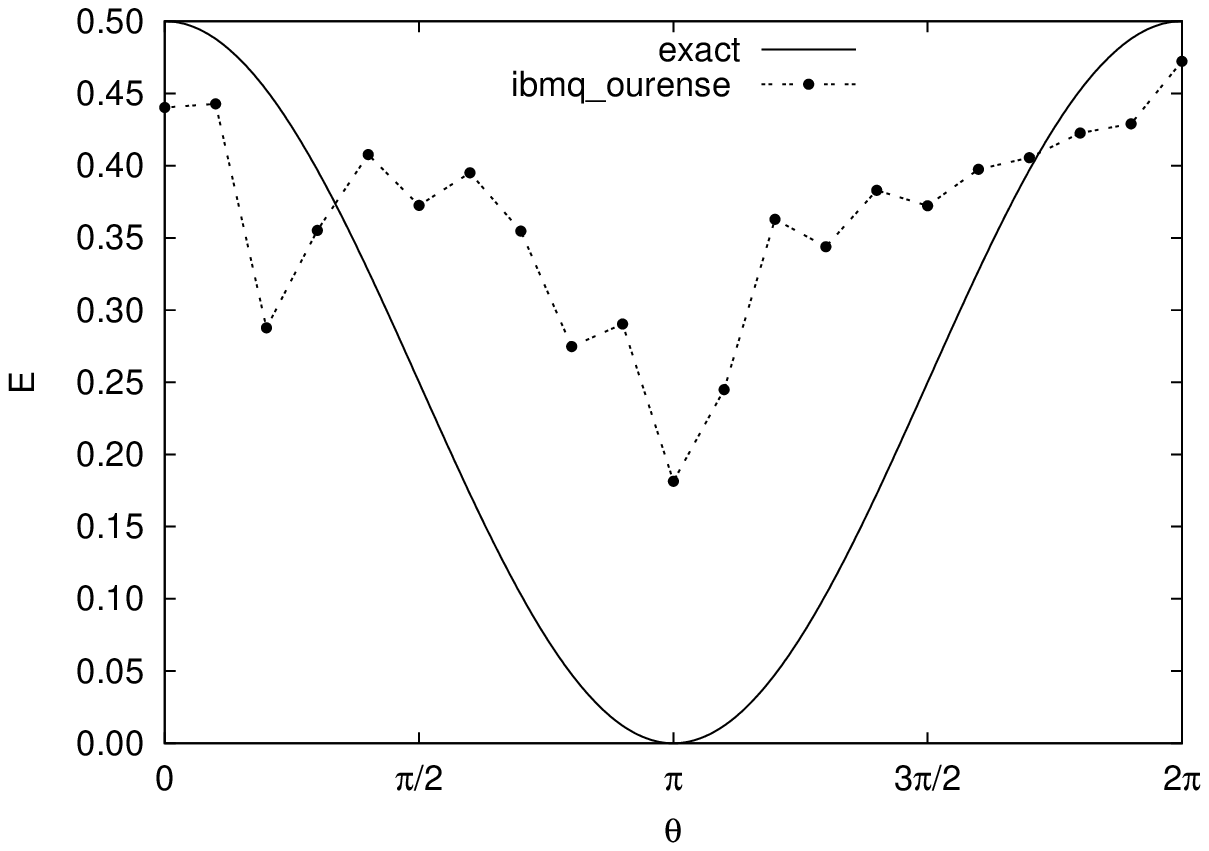}}
\subcaptionbox{\label{EntWq4}}{\includegraphics[scale=0.30, angle=0.0, clip]{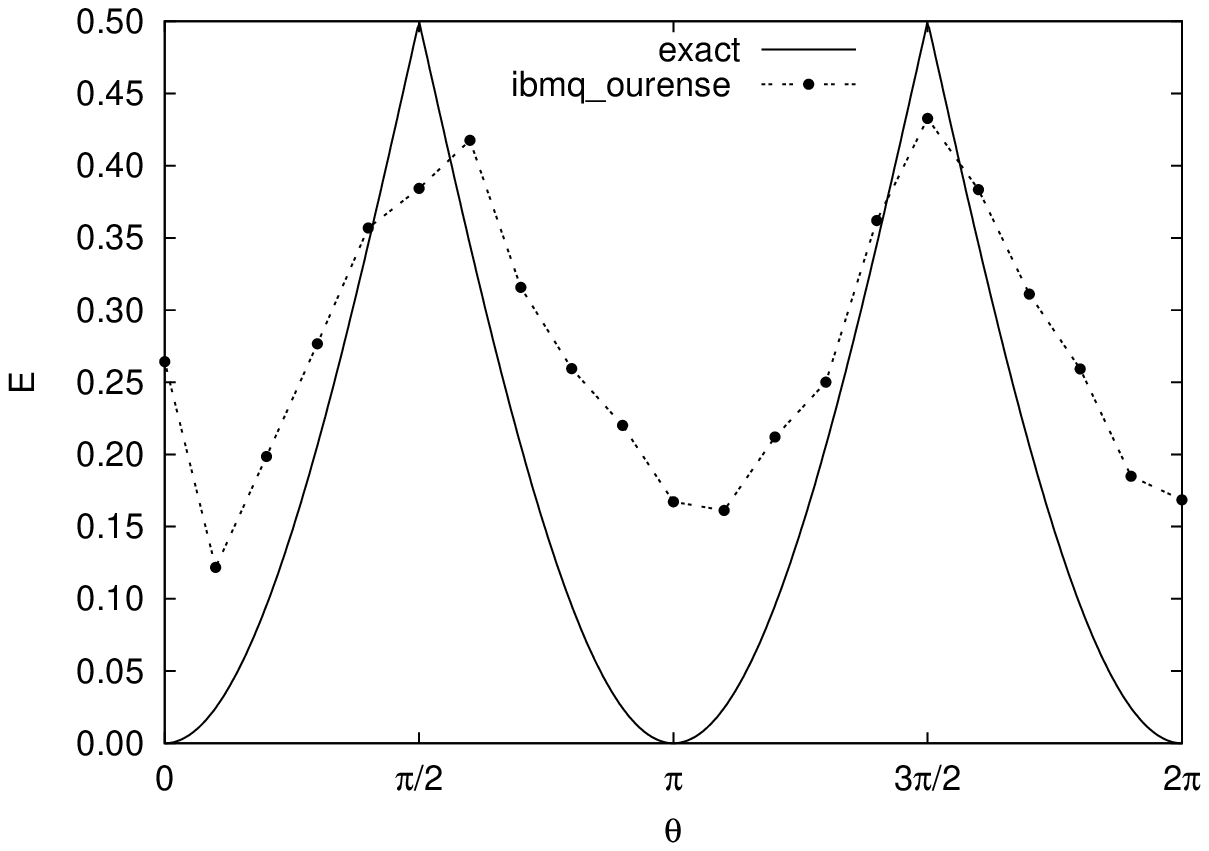}}
\caption{Dependence of the value of entanglement of the Werner state (\ref{Wernerstate}) on parameter $\theta$ in the cases of measuring the mean values
of q[1] (a), q[3] (b) and q[4] (c) spins.}
\label{EntWS}
\end{figure}

Because of the asymmetric nature of state (\ref{Wernerstate}), the expression which defines the entanglement of the first or second qubit with other qubits
is different from the expression which defines the entanglement of the third qubit with the first two. This expressions have the following form
\begin{eqnarray}
&&E_{1,2}\left(\vert\psi_{W}\rangle\right)=\frac{1}{2}\cos^2\frac{\theta}{2},\nonumber\\
&&E_3\left(\vert\psi_{W}\rangle\right)=\frac{1}{2}\left(1-\vert\cos\theta\vert\right).
\label{geommeasure1}
\end{eqnarray}
The results of measuring the values of entanglement of each qubit with the two others by the ibmq-ourense quantum device are presented in Fig.~\ref{EntWS}.
In this case, the results obtained by the quantum computer are worse than the theoretical ones. This is because the number of basis operators which should be applied to the system to prepare the Werner state reduces the accuracy of achievement of this state, which in turn affects the value of entanglement. The controlled-NOT operator of circuit presented in Fig.~\ref{Wstate} connects q[1] and q[4] qubits separated by the q[3] one (see Fig.~\ref{ibmq-ourense}), and this operator cannot be realized directly on the ibmq-ourense quantum computer. Therefore, the quantum computer uses the transpiled circuit. Instead of the q[1], q[3] and q[4] qubits, it uses the q[0], q[1] and q[3] qubits. For this new set of qubits, the Toffoli gate is presented by a set of basis operators which consists of 15 controlled-NOT and 8 single-qubit operators. Thus, the transpiled scheme uses 18 controlled-NOT and 13 single-qubit operators. Such a large number of gates leads to the accumulation of errors.

\section{Measuring the entanglement of mixed states prepared on the ibmq-ourense quantum computer \label{sec4}}

In paper \cite{frydryszak2017} the geometry measure of entanglement of rank-2 mixed states by correlations between qubits was studied.
The authors obtained an expression which allows one to calculate the entanglement of any qubit with another one in the mixed state defined
by the density matrix $\rho=\sum_{\alpha} \omega_{\alpha}\vert\psi_{\alpha}\rangle\langle\psi_{\alpha}\vert$, where vectors $\vert\psi_{\alpha}\rangle$
are given on the subspace spanned by vectors $\vert{\bf 0}\rangle=\vert 00\ldots 0\rangle$, $\vert{\bf 1}\rangle=\vert 11\ldots 1\rangle$,
and $\sum_{\alpha} \omega_{\alpha}=1$. In the case of an $n$-qubit state, the value of entanglement of certain qubit denoted by $i$ with another qubits
is determined by the expression
\begin{eqnarray}
E\left(\rho\right)=\frac{1}{2}\left(1-\sqrt{1-\langle\Sigma^x\rangle^2-\langle\Sigma^y\rangle^2}\right).
\label{mixedstateent}
\end{eqnarray}
The operators $\Sigma^x=\sigma_1^x\sigma_2^x\ldots\sigma_i^x\ldots\sigma_n^x$, $\Sigma^y=\sigma_1^x\sigma_2^x\ldots\sigma_i^y\ldots\sigma_n^x$, $\Sigma^z=I_1I_2\ldots\sigma_j^z\ldots I_n$
are the analogs of the Pauli operators acting on the subspace spanned by $\vert{\bf 0}\rangle$, $\vert{\bf 1}\rangle$.
Here $I_i$ is the unity single-spin operator and $\sigma_j^z$ is dependent on the qubit number $j$. Note that the
analog of the Pauli operators for any two-dimensional Hilbert subspace can be introduced in a similar way. Then the value of entanglement of a 2-rank mixed state is defined by expression (\ref{mixedstateent}) with the Pauli operators defined on this subspace.

\begin{figure}[!!h]
\subcaptionbox{\label{mixed_Bell1}}{\includegraphics[scale=0.44, angle=0.0, clip]{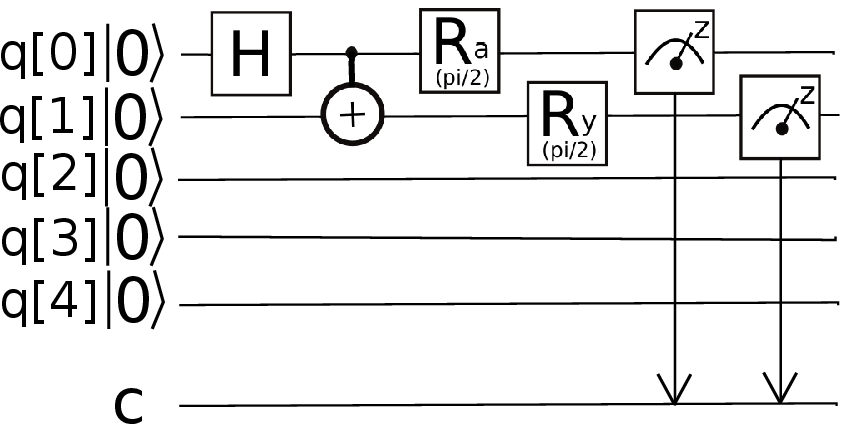}}
\subcaptionbox{\label{mixed_Bell2}}{\includegraphics[scale=0.44, angle=0.0, clip]{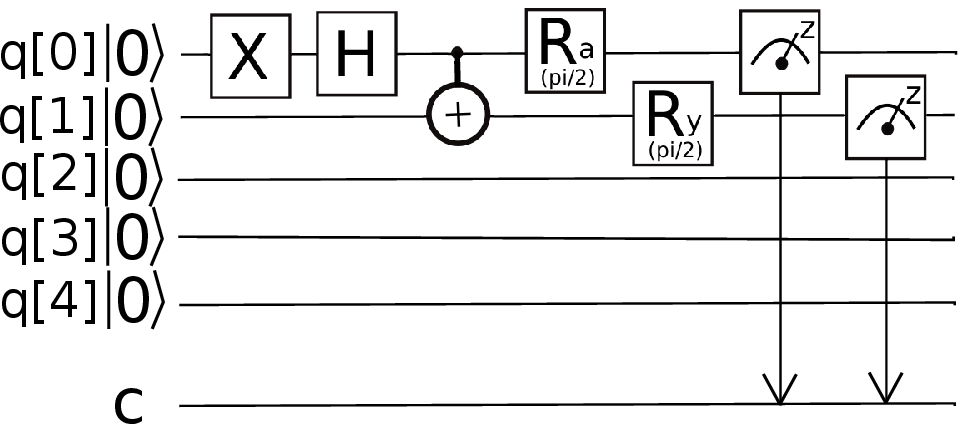}}\\
\caption{Circuits for preparation and measurement of the correlations of $\vert\Phi^+\rangle$ (a)
and $\vert\Phi^-\rangle$ (b) Bell states. Operators $R_a$ provide the rotations of the
qubit state around the $a=x,y$ axises by an angle $\pi/2$.}
\label{mixed_Bell}
\end{figure}

\begin{figure}[!!h]
\subcaptionbox{\label{mixed_state_ourense}}{\includegraphics[scale=0.33, angle=0.0, clip]{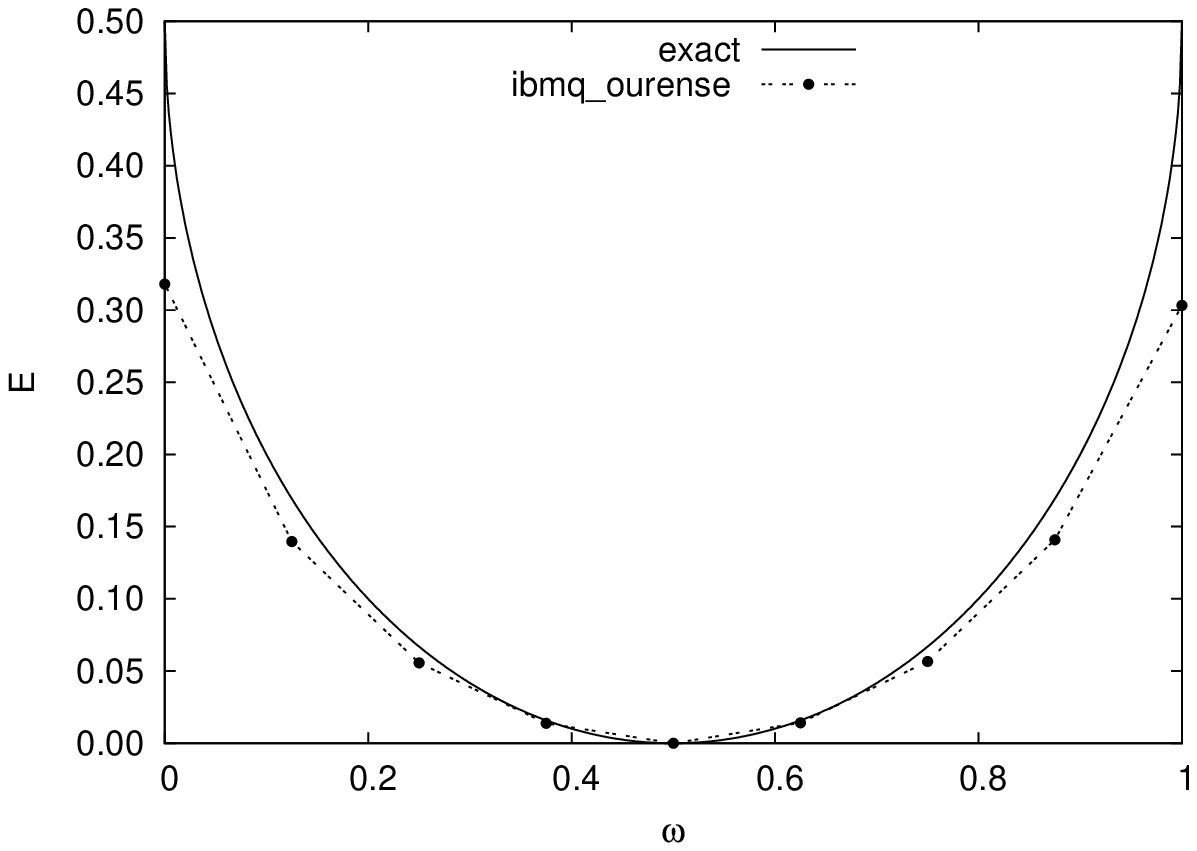}}
\subcaptionbox{\label{mixed_state_ourense_dev}}{\includegraphics[scale=0.33, angle=0.0, clip]{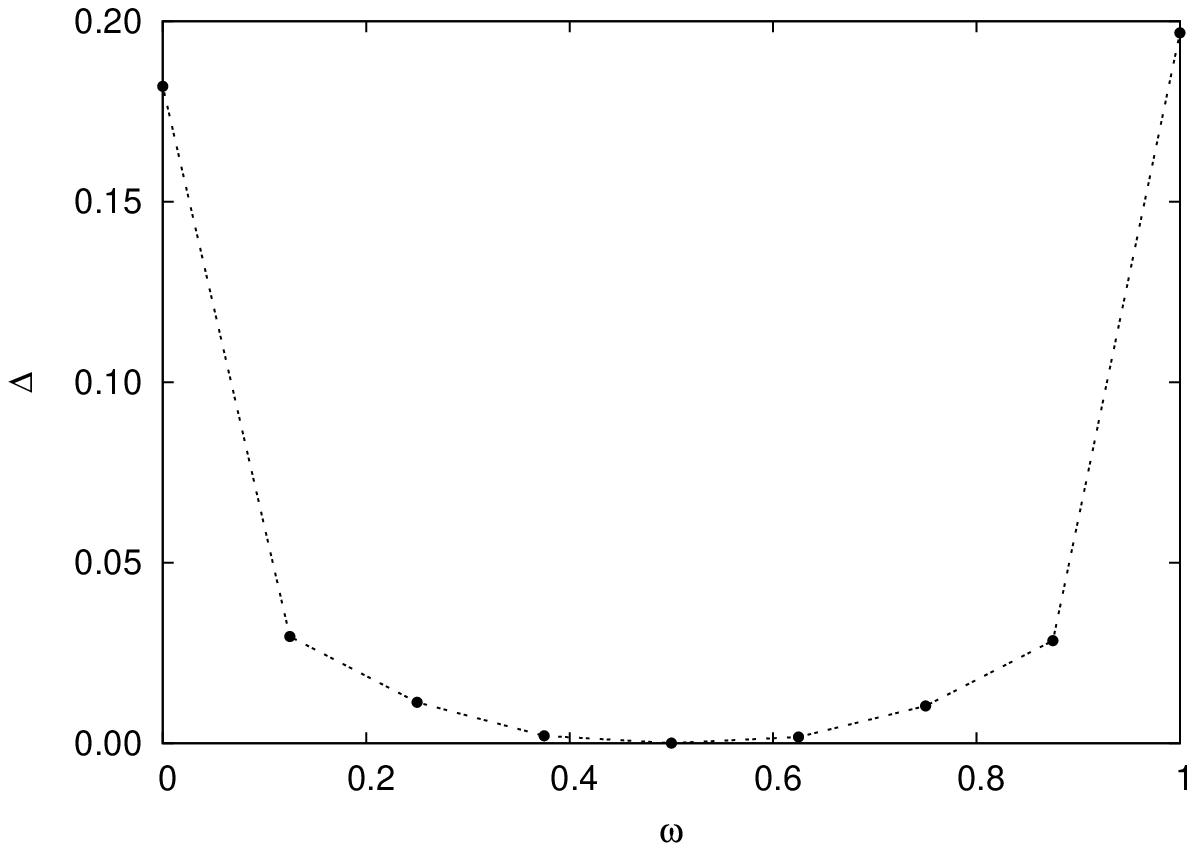}}
\caption{Dependence of (a) entanglement of state (\ref{mixedBellstate}) and (b) difference between the theoretical
and measured values of this entanglement on the mixing parameter are depicted.}
\label{mixed_Bell}
\end{figure}

Similarly, as in the case of a pure state, the correlation functions in equation (\ref{mixedstateent}) can be expressed by probabilities.
However, unlike the previous case, where only the one spin is measured, here the entire system should be measured. In the case of a two-qubit state, the correlation functions in equation (\ref{mixedstateent}) take the form
\begin{eqnarray}
\langle\sigma_1^a\sigma_2^x\rangle = \sum_{\alpha}\omega_{\alpha}\langle\psi_{\alpha}\vert\sigma_1^a\sigma_2^x\vert\psi_{\alpha}\rangle,
\label{corrfuncx}
\end{eqnarray}
where $a=x,y$. This expression contains the correlation functions constructed on pure states $\vert\psi_{\alpha}\rangle$. These functions can be expressed by the probabilities measured in the experiment as follows
\begin{eqnarray}
&&\langle\psi_{\alpha}\vert\sigma_1^x\sigma_2^x\vert\psi_{\alpha}\rangle\nonumber\\
&&=\vert\langle{\tilde \psi}_{\alpha}^{yy}\vert 00\rangle\vert^2-\vert\langle{\tilde \psi}_{\alpha}^{yy}\vert 01\rangle\vert^2-\vert\langle{\tilde \psi}_{\alpha}^{yy}\vert 10\rangle\vert^2+\vert\langle{\tilde \psi}_{\alpha}^{yy}\vert 11\rangle\vert^2,\nonumber\\
&&\langle\psi_{\alpha}\vert\sigma_1^y\sigma_2^x\vert\psi_{\alpha}\rangle\nonumber\\
&&=\vert\langle{\tilde \psi}_{\alpha}^{xy}\vert 00\rangle\vert^2-\vert\langle{\tilde \psi}_{\alpha}^{xy}\vert 01\rangle\vert^2-\vert\langle{\tilde \psi}_{\alpha}^{xy}\vert 10\rangle\vert^2+\vert\langle{\tilde \psi}_{\alpha}^{xy}\vert 11\rangle\vert^2,\nonumber\\
\label{mvcorrfunc}
\end{eqnarray}
where $\vert{\tilde \psi}_{\alpha}^{yy}\rangle=e^{i\frac{\pi}{4}\left(\sigma_1^y+\sigma_2^y\right)}\vert\psi_{\alpha}\rangle$, $\vert\tilde{\psi}^{xy}\rangle=e^{-i\frac{\pi}{4}\left(\sigma_1^x-\sigma_2^y\right)}\vert\psi\rangle$.
As an example, we consider a mixed state $\rho_{Bell}$ constructed of two Bell states $\vert\Phi^{\pm}\rangle=1/\sqrt{2}(\vert 00\rangle\pm\vert 11\rangle)$ as follows
\begin{eqnarray}
\rho_{Bell}=\omega\vert\Phi^+\rangle\langle\Phi^+\vert+\left(1-\omega\right)\vert\Phi^-\rangle\langle\Phi^-\vert,
\label{mixedBellstate}
\end{eqnarray}
where $\omega\in[0,1]$. The exact expression of geometric measure of entanglement of this system according to (\ref{mixedstateent}) is
\begin{eqnarray}
E\left(\rho_{Bell}\right)=\frac{1}{2}\left(1-2\sqrt{\omega(1-\omega)}\right).
\label{entmixedBellstate}
\end{eqnarray}
On the quantum device ibmq-ourense, we separately prepare and measure the correlations of the $\vert\Phi^+\rangle$ (Fig.~\ref{mixed_Bell1}) and
$\vert\Phi^-\rangle$ (Fig.~\ref{mixed_Bell2}). Note that the number of measurements of pure states should be made according to their weights defined by $\omega$. Namely, to find the entanglement of state (\ref{mixedBellstate}), we measure the states  $\vert\Phi^+\rangle$
and $\vert\Phi^-\rangle$ as shown in Fig.~\ref{mixed_Bell1} and Fig.~\ref{mixed_Bell2}, respectively. Similarly as in the previous case, to obtain the probabilities which define the mean values (\ref{mvcorrfunc}), the quantum computer makes some number of shots. As we mentioned, this number of shots should be proportional to the weights which are in state (\ref{mixedBellstate}). For instance, in the case of the maximally mixed state ($\omega=1/2$) the numbers of shots are the same for both states. Thus, for predefined weights the quantum computer makes certain numbers of necessary measurements in $\vert\Phi^+\rangle$ and $\vert\Phi^-\rangle$ states. However, for the states with different weights, the total number of measurements is fixed. In our case, the total number of measurements is equal to 8192. The numbers of measurements, which should be made by the quantum computer for the predefined mixed state (\ref{mixedBellstate}), are defined as follows: $8192\times\omega$ for $\vert\Phi^+\rangle$ state and $8192\times(1-\omega)$ for $\vert\Phi^-\rangle$ one. We prepare the states with $\omega$ which changes with step 0.125. Then, the number of shots is changed with step 1024.
The exact behavior (\ref{entmixedBellstate}) and behavior of entanglement obtained by a quantum computer with different value of $\omega$
are presented in Fig.~\ref{mixed_state_ourense}. The difference between theoretical ($E$) and  measured ($E_i$)
values of entanglement, $\Delta=\vert E_i-E\vert$, for certain $\omega$ is depicted in Fig.~\ref{mixed_state_ourense_dev}. As we can see,
for more mixed states the entanglement is determined more precisely. It should be noted that the same trend is observed
when determining the relative deviation $\delta=\vert E_i-E\vert /E$. In our opinion, this is because the maximally
mixed state (with $\omega=1/2$) (\ref{mixedBellstate}) can be also written in the form $1/2(\vert 00\rangle\langle 00\vert+\vert 11\rangle\langle 11\vert)$ which is spanned by the basis vectors $\vert 00\rangle$, $\vert 11\rangle$. These vectors are both the eigenstates of the density matrix and the basis states on which the quantum computer makes the measurements. It should be stressed once again that the eigenstates of the density matrix are the Bell states $\vert\Phi^+\rangle$ and $\vert\Phi^-\rangle$. However, in the case of $\omega=1/2$ the eigenvalues become degenerate and the states $\vert 00\rangle$, $\vert 11\rangle$ are also the eigenstates of the density matrix. Therefore, we think that coincidence of eigenstates of density matrix and basis states of the quantum computer provides a good agreement between the experimental value and theoretical prediction in the case of $\omega=1/2$. Thus, the closer the state is to the maximally mixed state, the more accurate it is measured.

\section{Conclusions}

We have considered the protocol which allows one to determine the value of entanglement between a qubit and the rest of the system prepared on a quantum computer for the pure and special mixed quantum states. This protocol implement results from paper \cite{frydryszak2017}. In the case of pure states  this protocol is based on the calculation of the mean value of spin. We have applied this protocol to states, namely, the Schr\"odinger cat and Werner-like states, by the ibmq-ourense quantum computer. In the case of the Schr\"odinger cat states, the results obtained on its are in good agreement with the theoretical ones. The error increases a little bit with respect to the number of qubits in the system.
This is because the gates that generate the Schr\"odinger cat state are basis operators of the ibmq-ourense quantum
device. Another situation we have in the case of three-qubit Werner states, where its preparation requires additional operators. This leads to an accumulation of errors and worse agreement with theoretical
predictions. It should be noted that the measurement of only one spin of the system allows to minimize the measurement error.
This fact allows one to determined the entanglement of pure state in good agreement with the theory. We have also generalized
the protocol to determine the value of entanglement of rank-2 mixed states. In this case the correlations of all spins of the system
should be measured. As an example, we have obtained the value of entanglement of the mixed state which consists of two Bell states
$\vert\Phi^{\pm}\rangle=1/\sqrt{2}(\vert 00\rangle\pm\vert 11\rangle)$ (see Fig.~\ref{mixed_state_ourense}). In this case we also obtained good agreement with the theoretical prediction. For more mixed states we obtain the smaller deviation from the theory.
Thus, more mixed states are measured more accurately. This is because the maximally mixed state (\ref{mixedBellstate})
is spanned by the basis vectors of quantum computer that reduces the measurement error.

\section{Acknowledgements}
We thank Profs. Andrij Rovenchak and Svyatoslav Kondrat (aka Valiska) for useful comments.
This work was supported by Project FF-83F (No.~0119U002203) from the Ministry of Education and Science of Ukraine.

\end{document}